\documentclass{cimento}

\usepackage{graphicx}
\usepackage[british]{babel}
\graphicspath{{figures/}}
\usepackage{amsmath,amssymb}

\title{In-plasma study of opacity relevant for compact binary ejecta }
\author{A.~Pidatella\from{A}\thanks{pidatella@lns.infn.it}\ETC, 
S.~Cristallo\from{B}\from{B1},
A.~Galat\`a\from{C},
M.~La Cognata\from{A},
M.~Mazzaglia\from{A},
A.~Perego\from{D}\from{D1},
R.~Spart\`a\from{A},
A.~Tumino\from{A}\from{E}, 
D.~Vescovi\from{F}\from{B1}, \atque
D.~Mascali\from{A}
}
\instlist{\inst{A} INFN, LNS - Catania, Italy
  \inst{B} INAF, Osservatorio Astronomico d'Abruzzo - Teramo, Italy
  \inst{B1} INFN, Sezione di Perugia - Perugia, Italy
  \inst{C} INFN, LNL - Legnaro, Italy
  \inst{D}Dipartimento di Fisica, Universit\`a di Trento - Trento, Italy
  \inst{D1} TIFPA, INFN - Trento, Italy
  \inst{E}Facolt\`a di Ingegneria e Architettura, Universit\`a degli Studi di Enna "Kore" - Enna, Italy
  \inst{F}Goethe University Frankfurt - Frankfurt am Main, Germany}

\begin{document}

\maketitle

\begin{abstract}
In the context of the INFN project \textbf{PANDORA{\_}Gr3} (\textbf{P}lasma for \textbf{A}strophysics, \textbf{N}uclear \textbf{D}ecays \textbf{O}bservation and \textbf{R}adiation for \textbf{A}rchaeometry) and of multi-messenger astronomy, we propose a feasibility study for in-laboratory plasma's opacity investigation, in an environment resembling thermodynamic conditions typical of the ejecta of compact binary mergers  containing at least a neutron star. We aim to advance knowledge on the physics of \emph{kilonovae}, the electromagnetic transients following a merger, which are relevant for the study of the origin of heavy nuclei in the Universe produced via \emph{r}-process nucleosynthesis. In this paper, we present preliminary results of numerical simulations for some physics cases considered in the light of a possible experimental setup for future in-laboratory opacity spectroscopic measurements. 
\end{abstract}

\section{Introduction}
Binary neutron star (BNS) and black hole-neutron star (BH-NS) mergers can expel rapidly expanding matter, usually referred to as compact binary ejecta (CBE). At initial stages, CBE are very rich in neutrons and have peculiar thermodynamic conditions which are compatible with being among the major sources of cosmic rapid neutron capture nucleosynthesis (\emph{r}-process)  \cite{Korobkin2012}. This matter, enriched in freshly synthesized radioactive isotopes, powers electromagnetic transient emissions known as \emph{kilonovae} \cite{Arcavi2017}. These signals (from UV to near-IR) represent one of the primary electromagnetic counterparts of gravitational-wave events produced by the merging, as it occurred for the GW170817 event \cite{Abbott2017}. A kilonova arises from a \emph{translucent} stage of the expanding ejecta, when thermal radiation can escape. Its energy results from a balance between thermalization processes and the radioactive warm-up  due to nuclear fissions and decays \cite{Li1998,Metzger2010}. 
As a result of the non-trivial merging dynamics, several ejection episodes can occur and, depending on the ejecta neutron richness, both \textit{heavy} and \textit{light} elements are synthesized via the \textit{r}-process. The presence of heavy and light \textit{r}-process elements can be disentangled by analysing the kilonova light curve. Previous studies of the bolometric and broad band light curves  of AT2017gfo (GW170817 kilonova) \cite{Arcavi2017} have stressed the presence of at least two ejecta's components (in terms of mass, velocity, and composition) \cite{Kasen2017}. 
Due to this complex composition, the wavelength, intensity, and evolution time-scale of the kilonova light curve are strongly affected by the ejecta \emph{opacity} \cite{Kasen2017,Barnes2013,Kasen2013}. Opacity, $\kappa(\nu)\sim N_{ij}\sigma^{ph}_{j\rightarrow k}(\nu)$, with $N_{ij}$ being the atomic level population of excited level $j$ of ion stage $i$, and $\sigma^{ph}_{j\rightarrow k}(\nu)$ being the cross section for the atomic transition from level $j$ to level $k$, regulates the energy exchange between radiation and plasma. It arises from the blending of millions of atomic line transitions, and considers multiple absorption-scattering processes involved in the radiation transport from its nucleation place to the edge of the ejecta. Here, opacity can differ of several orders of magnitudes: ejecta enriched in light \textit{r}-process elements have relatively low opacity ($\kappa\lesssim 1$ cm g$^{-1}$), radiating \textit{optical} light that fades in days, while heavy \textit{r}-process elements enlarges the opacity ($\kappa\approx 10$ cm g$^{-1}$), with \textit{redder} light curves lasting even for weeks, depending on the complexity of contributing atomic sub-shells \cite{Kasen2017}.
State-of-the-art works point to the necessity to make progress in opacity modelling \cite{Kasen2013,Barnes2016,Tanaka2019} for more correct kilonova light curve predictions. Available models of these objects are often oversimplified and can result incomplete or inconsistent when compared with observational data.
A more detailed knowledge of kilonovae events is important both for astrophysics and nuclear astrophysics to quickly identify related gravitational-wave events, and also to draw quantitative conclusions on the \textit{r}-process nucleosynthetic yield from observations, relevant in the study of the cosmic origin of \emph{r}-process nuclei.

In the context of the INFN project PANDORA{\_}Gr3 \cite{Mascali2017}, we propose a feasibility study of CBE plasma opacity evaluation. We aim at designing an experimental setup for in-laboratory opacity measurements, relevant for kilonovae light curve predictions. In the following, we present preliminary results of numerical simulations aiming at locating reproducible plasma conditions of CBE in the laboratory, and estimates of their nuclei abundances according to the \emph{r}-process nucleosynthesis. These studies permit to select some remarkable physical cases, for which a systematic numerical spectral analysis has been carried out. This investigation has provided stronger bases to further proceed with the experimental design of opacity measurements.

\section{Modelling of kilonova-emitting CBE and \emph{r}-process nucleosynthetic yields}
\begin{figure}
\centering
  \includegraphics[width=\textwidth]{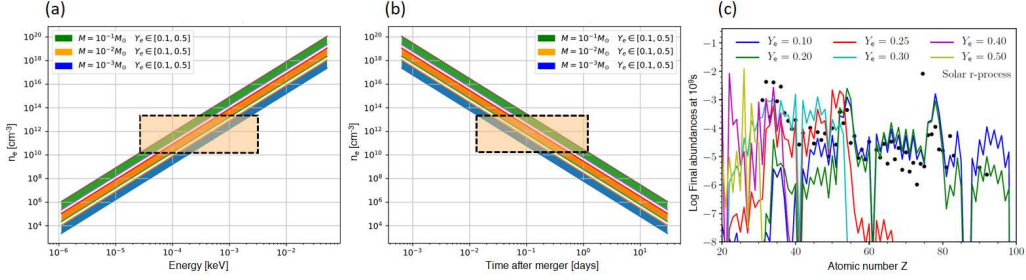}
  \caption{(\textit{a}) Evolution of ejecta parameters in the electron density $n_e \,[\mathrm{cm^{-3}}]$~\textit{vs.}~energy [\un{keV}] (computed as $k_B T$) plane, for ejecta moving with velocity $v=0.3c$. (\textit{b}) Time-evolution of electron density of ejecta adiabatically expanding at $v=0.3c$. Different colors refer to different ejecta masses $M$ and initial electron fraction $Y_e$. Dashed coloured boxes locate \textit{laboratory} feasibility conditions. (\textit{c}) Elements abundances for \textit{r}-process nucleosynthesis assuming different $Y_e$ as calculated with the SKYNET \cite{Lippuner2017} nuclear network. } 
  \label{ejecta}
\end{figure}

Inside the PANDORA{\_}Gr3 compact magnetic trap, a dense and hot plasma, made of multi-charged ions in a cloud of energetic electrons is confined in a so-called minimum-B magnetic profile, and heated by microwave power, according to the electron cyclotron resonance (ECR) mechanism \cite{Mascali2017}. The generated ECR plasmas can reach electron densities up to $\sim 10^{13} \mathrm{cm^{-3}}$ and energies in the range of $1~\mathrm{eV} - 10~\mathrm{keV}$. As first step of this study, we have located the CBE stage at which plasma conditions are reproducible in laboratory plasma. For this purpose, modelling of time evolution of CBE has been carried out considering homologous expansion of a fluid element under adiabatic conditions \cite{Korobkin2012}. Initial conditions for the \emph{mass} $M$, \emph{temperature} $T$, \emph{velocity} $v$, and \emph{electron fraction} $Y_e$ - the latter indicative of the initial neutron richness - were assumed from the state-of-the-art numerical simulations outcome for BNS mergers \cite{Martin2015,Radice2018}. Some results are shown in figs.~\ref{ejecta}(\textit{a,b}).
The expected evolution in the density-energy plane in fig.~\ref{ejecta}(\textit{a}) suggests that ejecta resemble ECR plasmas densities in an energy range of a few eV. These plasma parameters, as suggested by fig.~\ref{ejecta}(\textit{b}), fit better the conditions of early-stage kilonova emission, \idest , between $10^{-2} - 10^{0}$ days after merger.
This early phase of the signal (\textit{blue-kilonova emission}) has its peak at optical frequencies, more likely due to the ejecta's \textit{light} component featuring a low degree of opacity. Assuming those plasma conditions for CBE, we have determined the main atomic abundances in the astrophysical environment, according to the \textit{r}-process nucleosynthesis, in order to constrain relevant elements for in-laboratory measurements.
The neutron richness of the ejecta plays a fundamental role in determining the nucleosynthesis outcome. In fig. \ref{ejecta}(\textit{c}) we report on several \textit{r}-process yields distributions computed by means of the SKYNET nuclear network \cite{Lippuner2017} as a function of $Y_e$. It can be evinced that light \textit{r}-process elements production dominates for $Y_e\geq 0.25$, where large $Y_e$'s are expected for early-days blue kilonova (being a reproducible experimental condition as shown in figs. \ref{ejecta}(\textit{a,b})). On this basis, and since the blue-kilonova emission is more likely shaped by light \textit{r}-process elements \cite{Tanaka2019,Kasen2013bis,Watson2019}, we have initially considered elements going from \textit{selenium} to \textit{rhodium} as eligible for our experimental campaign.

\section{Numerical simulations about opacity in PANDORA{\_}Gr3 plasmas}

\begin{figure}
\centering
  \includegraphics[width=\textwidth]{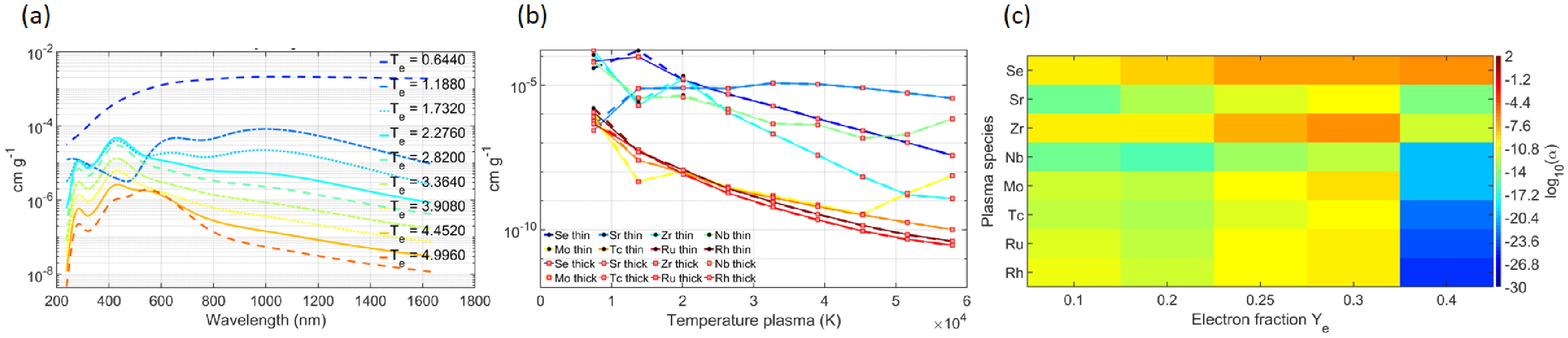}
  \caption{(\textit{a}) Opacity ($\mathrm{cm~g^{-1}}$) \textit{vs.} wavelength ($\mathrm{nm}$) for NLTE optically thick \textit{selenium} plasma, at density $\rho=10^{12}~\mathrm{cm^{-3}}$ and various energies $k_B T$\un{(eV)}. (\textit{b}) Frequency-integrated mean opacity ($\mathrm{cm~g^{-1}}$) \textit{vs.} plasma temperature $T$\un{(K)}, for all single-species NLTE plasmas, at $\rho=10^{12}~\mathrm{cm^{-3}}$. Optical thin (dashed-line, circle marker) and thick (solid-line, square marker) cases are shown. (\textit{c}) Weighted opacity $\log(\alpha)$, in the atomic elements - $Y_e$ plane, for NLTE optically-thick plasmas, at $\rho=10^{12}~\mathrm{cm^{-3}}$ and energy $k_{B}T=0.644$~\un{eV}. } 
  \label{flychk}
\end{figure}

In addition to considering the abundances of light \textit{r}-process elements in CBE corresponding to early-epochs kilonova emission, we have also based the choice of plasma species for experiments on their contribution to opacity. For this purpose, a campaign of numerical simulations to explore the expected opacity contribution in the \textit{visible} (VIS) range has been performed by using FLYCHK \cite{Chung2005}, a population kinetics and spectral modelling code that can evaluate plasma opacity, according to the degree of plasma ionization and atomic level population distribution. These aspects strongly depend on the plasma parameters. The code accounts both for finite-size plasma and different plasma models, such as local thermodynamic equilibrium (LTE) and non-LTE (NLTE), to produce theoretical synthetic spectra, given plasma density $\rho$, energy $k_{B}T$, and \textit{optical thickness} $\tau$. As a first study, single-species self-emitting plasmas made of Se, Sr, Zr, Nb, Mo, Tc, Ru, and Rh have been considered. Both LTE and NLTE regimes were investigated, for $\rho \sim 10^{9} \div 10^{14}~\mathrm{cm^{-3}}$, $k_{B}T \sim 0.6 \div 5~\mathrm{eV}$, and $\tau = 1~\mu\mathrm{m},10~\mathrm{cm}$. Some of the numerical results are shown in figs. \ref{flychk}(\emph{a-c}).
Figure \ref{flychk}(\textit{a}) shows an example of opacity spectrum from simulations for NLTE optically thick Se plasma, at fixed density and various energies. Several contributions to opacity in the VIS are present. Comparison with the LTE counterpart has evinced a better resolution of the self-consistent interplay between rate and radiative transport equations, whilst thicker plasma contributes with larger opacities than thin plasmas. To describe the plasma's tendency in absorbing radiation in the studied spectral range, frequency-integrated opacity, \idest, \textit{mean opacity}, has been evaluated. An example of mean opacity calculations is shown in fig. \ref{flychk}(\textit{b}), for all single-species plasmas. There is a small group of few elements, \idest, Se, Sr, Zr, and Nb, exhibiting larger mean opacity at the temperature condition of early-epochs kilonova ($\lesssim 2\cdot 10^4 $\un{K}). In view of these numerical results, with the purpose of selecting the most suitable species for our experimental measurements, we have defined a \textit{weighted opacity} parameter $\alpha$, \idest, species' mean opacity weighted by their abundances at a given $Y_e$, as given by SKYNET results shown in fig. \ref{ejecta}(\textit{c}). Results shown in fig. \ref{flychk}(\textit{c}) suggest, for $Y_e\geq0.25$ and $T$ typical of blue-kilonova emission, selenium plasma as one of the most favoured for the experiment.
\section{Conclusion}
The presented study has provided bases for future experimental activity on plasma opacity for kilonova light curve predictions. It has especially given constraints in terms of plasma species eligible for being studied in the laboratory, plasma conditions, spectral range and expected features. This information turns out to be useful for optimizing the experimental setup. One important aspect is the strong impact of NLTE thermodynamic conditions on plasma spectral features, where the astrophysical scenario is often assumed in LTE \cite{Tanaka2019,Kasen2013bis,Pinto2000}. While theoretical expectations based only on \textit{r}-process abundances and number of transition lines make elements as Mo, Tc, Ru, and Rh the most opaques among those considered \cite{Tanaka2019}, their resulting mean opacities are several orders of magnitude lower than those for Se or Sr NLTE plasma. Thus, thermodynamic conditions can play an important role in determining opacity of elements. To conclude, further numerical investigations are planned, in view of future measurements, exploring  the impact of multiple-species plasma and external radiation field on the plasma spectra.

\acknowledgments
The authors wish to thank support of INFN through the project PANDORA{\_}Gr3 funded by 3rd Nat. Sci. Comm..

\end{document}